\documentclass[prl,
secnumarabic,twocolumn,twoside,a4paper,nofootinbib,nobibnotes,showpacs,amsmath,amssymb,floatfix%
]{revtex4}

\newcommand{\usetxfonts}{\usepackage{txfonts}}

\usepackage{savesym} 
\usepackage[T1]{fontenc} 
\usepackage{textcomp}
\usepackage{amsxtra}

\usepackage[british]{babel} 

\savesymbol{qedsymbol}

\usepackage{amsthm}

\restoresymbol{ams}{qedsymbol}
\newcommand{\QEE}
{\amsqedsymbol}

\usepackage[dvipdf]{graphicx}
\usepackage{url}

\providecommand{\usetxfonts}{}
\usetxfonts

\DeclareFontFamily{U}{egreek}{\skewchar\font'177}%
\DeclareFontShape{U}{egreek}{m}{n}{<-6>s*[1]eurm5<6-8>s*[1]eurm7 <8->s*[1]eurm10}{}%
\DeclareFontShape{U}{egreek}{m}{it}{<->s*[1]eurmo10}{}%
\DeclareFontShape{U}{egreek}{b}{n}{<-6>s*[1]eurb5 <6-8>s*[1]eurb7 <8->s*[1]eurb10}{}%
\DeclareFontShape{U}{egreek}{b}{it}{<->s*[1]eurbo10}{}

\DeclareFontFamily{U}{egreek}{\skewchar\font'177}%
\DeclareFontShape{U}{egreek}{m}{n}{<-6>s*[1]eurm5<6-8>s*[1]eurm7 <8->s*[1]eurm10}{}%
\DeclareFontShape{U}{egreek}{m}{it}{<->s*[1]eurmo10}{}%
\DeclareFontShape{U}{egreek}{b}{n}{<-6>s*[1]eurb5 <6-8>s*[1]eurb7 <8->s*[1]eurb10}{}%
\DeclareFontShape{U}{egreek}{b}{it}{<->s*[1]eurbo10}{}%
\DeclareSymbolFont{egreeki}{U}{egreek}{m}{it}%
\SetSymbolFont{egreeki}{bold}{U}{egreek}{b}{it}
\DeclareSymbolFontAlphabet{\mathegri}{egreeki}%
\DeclareMathSymbol{\epartial}{\mathalpha}{egreeki}{"40}

\DeclareMathSymbol{\ealpha}{\mathalpha}{egreeki}{"0B}
\DeclareMathSymbol{\ebeta}{\mathalpha}{egreeki}{"0C}
\DeclareMathSymbol{\egamma}{\mathalpha}{egreeki}{"0D}
\DeclareMathSymbol{\edelta}{\mathalpha}{egreeki}{"0E}
\DeclareMathSymbol{\eepsilon}{\mathalpha}{egreeki}{"0F}
\DeclareMathSymbol{\ezeta}{\mathalpha}{egreeki}{"10}
\DeclareMathSymbol{\eeta}{\mathalpha}{egreeki}{"11}
\DeclareMathSymbol{\etheta}{\mathalpha}{egreeki}{"12}
\DeclareMathSymbol{\eiota}{\mathalpha}{egreeki}{"13}
\DeclareMathSymbol{\ekappa}{\mathalpha}{egreeki}{"14}
\DeclareMathSymbol{\elambda}{\mathalpha}{egreeki}{"15}
\DeclareMathSymbol{\emu}{\mathalpha}{egreeki}{"16}
\DeclareMathSymbol{\enu}{\mathalpha}{egreeki}{"17}
\DeclareMathSymbol{\exi}{\mathalpha}{egreeki}{"18}
\DeclareMathSymbol{\eomicron}{\mathalpha}{egreeki}{"6F}
\DeclareMathSymbol{\epi}{\mathalpha}{egreeki}{"19}
\DeclareMathSymbol{\erho}{\mathalpha}{egreeki}{"1A}
\DeclareMathSymbol{\esigma}{\mathalpha}{egreeki}{"1B}
\DeclareMathSymbol{\etau}{\mathalpha}{egreeki}{"1C}
\DeclareMathSymbol{\eupsilon}{\mathalpha}{egreeki}{"1D}
\DeclareMathSymbol{\ephi}{\mathalpha}{egreeki}{"1E}
\DeclareMathSymbol{\echi}{\mathalpha}{egreeki}{"1F}
\DeclareMathSymbol{\epsi}{\mathalpha}{egreeki}{"20}
\DeclareMathSymbol{\eomega}{\mathalpha}{egreeki}{"21}
\DeclareMathSymbol{\evarepsilon}{\mathalpha}{egreeki}{"22}
\DeclareMathSymbol{\evartheta}{\mathalpha}{egreeki}{"23}
\DeclareMathSymbol{\evarpi}{\mathalpha}{egreeki}{"24}
\let\evarrho\erho 
\let\evarsigma\esigma
\let\evarkappa\ekappa
\DeclareMathSymbol{\evarphi}{\mathalpha}{egreeki}{"27}
\DeclareMathSymbol{\evarAlpha}{\mathalpha}{egreeki}{"41}
\DeclareMathSymbol{\evarBeta}{\mathalpha}{egreeki}{"42}
\DeclareMathSymbol{\evarGamma}{\mathalpha}{egreeki}{"00}
\DeclareMathSymbol{\evarDelta}{\mathalpha}{egreeki}{"01}
\DeclareMathSymbol{\evarEpsilon}{\mathalpha}{egreeki}{"45}
\DeclareMathSymbol{\evarZeta}{\mathalpha}{egreeki}{"5A}
\DeclareMathSymbol{\evarEta}{\mathalpha}{egreeki}{"48}
\DeclareMathSymbol{\evarTheta}{\mathalpha}{egreeki}{"02}
\DeclareMathSymbol{\evarIota}{\mathalpha}{egreeki}{"49}
\DeclareMathSymbol{\evarKappa}{\mathalpha}{egreeki}{"4B}
\DeclareMathSymbol{\evarLambda}{\mathalpha}{egreeki}{"03}
\DeclareMathSymbol{\evarMu}{\mathalpha}{egreeki}{"4D}
\DeclareMathSymbol{\evarNu}{\mathalpha}{egreeki}{"4E}
\DeclareMathSymbol{\evarXi}{\mathalpha}{egreeki}{"04}
\DeclareMathSymbol{\evarOmicron}{\mathalpha}{egreeki}{"4F}
\DeclareMathSymbol{\evarPi}{\mathalpha}{egreeki}{"05}
\DeclareMathSymbol{\evarRho}{\mathalpha}{egreeki}{"50}
\DeclareMathSymbol{\evarSigma}{\mathalpha}{egreeki}{"06}
\DeclareMathSymbol{\evarTau}{\mathalpha}{egreeki}{"54}
\DeclareMathSymbol{\evarUpsilon}{\mathalpha}{egreeki}{"07}
\DeclareMathSymbol{\evarPhi}{\mathalpha}{egreeki}{"08}
\DeclareMathSymbol{\evarChi}{\mathalpha}{egreeki}{"58}
\DeclareMathSymbol{\evarPsi}{\mathalpha}{egreeki}{"09}
\DeclareMathSymbol{\evarOmega}{\mathalpha}{egreeki}{"0A} 
\DeclareFontFamily{U}{egreek}{\skewchar\font'177}%
\DeclareFontShape{U}{egreek}{m}{n}{<-6>s*[1]eurm5<6-8>s*[1]eurm7 <8->s*[1]eurm10}{}%
\DeclareFontShape{U}{egreek}{m}{it}{<->s*[1]eurmo10}{}%
\DeclareFontShape{U}{egreek}{b}{n}{<-6>s*[1]eurb5 <6-8>s*[1]eurb7 <8->s*[1]eurb10}{}%
\DeclareFontShape{U}{egreek}{b}{it}{<->s*[1]eurbo10}{}%
\DeclareSymbolFont{egreekr}{U}{egreek}{m}{n}%
\SetSymbolFont{egreekr}{bold}{U}{egreek}{b}{n}
\DeclareSymbolFontAlphabet{\mathegrr}{egreekr}%
\DeclareMathSymbol{\epartialup}{\mathalpha}{egreekr}{"40}

\DeclareMathSymbol{\ealphaup}{\mathalpha}{egreekr}{"0B}
\DeclareMathSymbol{\ebetaup}{\mathalpha}{egreekr}{"0C}
\DeclareMathSymbol{\egammaup}{\mathalpha}{egreekr}{"0D}
\DeclareMathSymbol{\edeltaup}{\mathalpha}{egreekr}{"0E}
\DeclareMathSymbol{\eepsilonup}{\mathalpha}{egreekr}{"0F}
\DeclareMathSymbol{\ezetaup}{\mathalpha}{egreekr}{"10}
\DeclareMathSymbol{\eetaup}{\mathalpha}{egreekr}{"11}
\DeclareMathSymbol{\ethetaup}{\mathalpha}{egreekr}{"12}
\DeclareMathSymbol{\eiotaup}{\mathalpha}{egreekr}{"13}
\DeclareMathSymbol{\ekappaup}{\mathalpha}{egreekr}{"14}
\DeclareMathSymbol{\elambdaup}{\mathalpha}{egreekr}{"15}
\DeclareMathSymbol{\emuup}{\mathalpha}{egreekr}{"16}
\DeclareMathSymbol{\enuup}{\mathalpha}{egreekr}{"17}
\DeclareMathSymbol{\exiup}{\mathalpha}{egreekr}{"18}
\DeclareMathSymbol{\eomicronup}{\mathalpha}{egreekr}{"6F}
\DeclareMathSymbol{\epiup}{\mathalpha}{egreekr}{"19}
\DeclareMathSymbol{\erhoup}{\mathalpha}{egreekr}{"1A}
\DeclareMathSymbol{\esigmaup}{\mathalpha}{egreekr}{"1B}
\DeclareMathSymbol{\etauup}{\mathalpha}{egreekr}{"1C}
\DeclareMathSymbol{\eupsilonup}{\mathalpha}{egreekr}{"1D}
\DeclareMathSymbol{\ephiup}{\mathalpha}{egreekr}{"1E}
\DeclareMathSymbol{\echiup}{\mathalpha}{egreekr}{"1F}
\DeclareMathSymbol{\epsiup}{\mathalpha}{egreekr}{"20}
\DeclareMathSymbol{\eomegaup}{\mathalpha}{egreekr}{"21}
\DeclareMathSymbol{\evarepsilonup}{\mathalpha}{egreekr}{"22}
\DeclareMathSymbol{\evarthetaup}{\mathalpha}{egreekr}{"23}
\DeclareMathSymbol{\evarpiup}{\mathalpha}{egreekr}{"24}
\let\evarrhoup\erhoup 
\let\evarsigmaup\esigmaup
\let\evarkappaup\ekappaup
\DeclareMathSymbol{\evarphiup}{\mathalpha}{egreekr}{"27}
\DeclareMathSymbol{\eAlpha}{\mathalpha}{egreekr}{"41}
\DeclareMathSymbol{\eBeta}{\mathalpha}{egreekr}{"42}
\DeclareMathSymbol{\eGamma}{\mathalpha}{egreekr}{"00}
\DeclareMathSymbol{\eDelta}{\mathalpha}{egreekr}{"01}
\DeclareMathSymbol{\eEpsilon}{\mathalpha}{egreekr}{"45}
\DeclareMathSymbol{\eZeta}{\mathalpha}{egreekr}{"5A}
\DeclareMathSymbol{\eEta}{\mathalpha}{egreekr}{"48}
\DeclareMathSymbol{\eTheta}{\mathalpha}{egreekr}{"02}
\DeclareMathSymbol{\eIota}{\mathalpha}{egreekr}{"49}
\DeclareMathSymbol{\eKappa}{\mathalpha}{egreekr}{"4B}
\DeclareMathSymbol{\eLambda}{\mathalpha}{egreekr}{"03}
\DeclareMathSymbol{\eMu}{\mathalpha}{egreekr}{"4D}
\DeclareMathSymbol{\eNu}{\mathalpha}{egreekr}{"4E}
\DeclareMathSymbol{\eXi}{\mathalpha}{egreekr}{"04}
\DeclareMathSymbol{\eOmicron}{\mathalpha}{egreekr}{"4F}
\DeclareMathSymbol{\ePi}{\mathalpha}{egreekr}{"05}
\DeclareMathSymbol{\eRho}{\mathalpha}{egreekr}{"50}
\DeclareMathSymbol{\eSigma}{\mathalpha}{egreekr}{"06}
\DeclareMathSymbol{\eTau}{\mathalpha}{egreekr}{"54}
\DeclareMathSymbol{\eUpsilon}{\mathalpha}{egreekr}{"07}
\DeclareMathSymbol{\ePhi}{\mathalpha}{egreekr}{"08}
\DeclareMathSymbol{\eChi}{\mathalpha}{egreekr}{"58}
\DeclareMathSymbol{\ePsi}{\mathalpha}{egreekr}{"09}
\DeclareMathSymbol{\eOmega}{\mathalpha}{egreekr}{"0A}

\newcommand{\alleugreek}{%
\let\alpha\ealpha
\let\beta\ebeta
\let\gamma\egamma
\let\delta\edelta
\let\epsilon\eepsilon
\let\zeta\ezeta
\let\eta\eeta
\let\theta\etheta
\let\iota\eiota
\let\kappa\ekappa
\let\lambda\elambda
\let\mu\emu
\let\nu\enu
\let\xi\exi
\let\omicron\eomicron
\let\pi\epi
\let\rho\erho
\let\sigma\esigma
\let\tau\etau
\let\upsilon\eupsilon
\let\phi\ephi
\let\chi\echi
\let\psi\epsi
\let\omega\eomega
\let\varepsilon\evarepsilon
\let\vartheta\evartheta
\let\varpi\evarpi
\let\varrho\evarrho 
\let\varsigma\evarsigma
\let\varkappa\evarkappa
\let\varphi\evarphi
\let\varAlpha\evarAlpha
\let\varBeta\evarBeta
\let\varGamma\evarGamma
\let\varDelta\evarDelta
\let\varEpsilon\evarEpsilon
\let\varZeta\evarZeta
\let\varEta\evarEta
\let\varTheta\evarTheta
\let\varIota\evarIota
\let\varKappa\evarKappa
\let\varLambda\evarLambda
\let\varMu\evarMu
\let\varNu\evarNu
\let\varXi\evarXi
\let\varOmicron\evarOmicron
\let\varPi\evarPi
\let\varRho\evarRho
\let\varSigma\evarSigma
\let\varTau\evarTau
\let\varUpsilon\evarUpsilon
\let\varPhi\evarPhi
\let\varChi\evarChi
\let\varPsi\evarPsi
\let\varOmega\evarOmega 
\let\alphaup\ealphaup
\let\betaup\ebetaup
\let\gammaup\egammaup
\let\deltaup\edeltaup
\let\epsilonup\eepsilonup
\let\zetaup\ezetaup
\let\etaup\eetaup
\let\thetaup\ethetaup
\let\iotaup\eiotaup
\let\kappaup\ekappaup
\let\lambdaup\elambdaup
\let\muup\emuup
\let\nuup\enuup
\let\xiup\exiup
\let\omicronup\eomicronup
\let\piup\epiup
\let\rhoup\erhoup
\let\sigmaup\esigmaup
\let\tauup\etauup
\let\upsilonup\eupsilonup
\let\phiup\ephiup
\let\chiup\echiup
\let\psiup\epsiup
\let\omegaup\eomegaup
\let\varepsilonup\evarepsilonup
\let\varthetaup\evarthetaup
\let\varpiup\evarpiup
\let\varrhoup\evarrhoup 
\let\varsigmaup\evarsigmaup
\let\varkappaup\evarkappaup
\let\varphiup\evarphiup
\let\Alpha\eAlpha
\let\Beta\eBeta
\let\Gamma\eGamma
\let\Delta\eDelta
\let\Epsilon\eEpsilon
\let\Zeta\eZeta
\let\Eta\eEta
\let\Theta\eTheta
\let\Iota\eIota
\let\Kappa\eKappa
\let\Lambda\eLambda
\let\Mu\eMu
\let\Nu\eNu
\let\Xi\eXi
\let\Omicron\eOmicron
\let\Pi\ePi
\let\Rho\eRho
\let\Sigma\eSigma
\let\Tau\eTau
\let\Upsilon\eUpsilon
\let\Phi\ePhi
\let\Chi\eChi
\let\Psi\ePsi
\let\Omega\eOmega
}
 
\usepackage{bm}
\providecommand{\piup}{\epiup}
\providecommand{\deltaup}{\edeltaup}
\providecommand{\zetaup}{\ezetaup}
\providecommand{\varepsilonup}{\evarepsilonup}

\providecommand{\coloneqq}{\mathrel{\mathop:}=}
\providecommand{\eqqcolon}{=\mathrel{\mathop:}}

\newcommand{\delt}{\deltaup}
\newcommand{\e}{\mathrm{e}}
\newcommand{\di}{\mathrm{d}}

\DeclareMathOperator{\tr}{tr}

\newcommand{\defd}{\coloneqq}
\newcommand{\defs}{\eqqcolon}

\newcommand{\cond}{\mathpunct{|}}
\newcommand{\with}{\colon}

\newcommand{\corr}{\mathrel{\hat{=}}}
\renewcommand{\ge}{\geqslant}
\DeclareMathDelimiter{\lclose}{\mathopen}{operators}{"5B}{largesymbols}{"02}
\DeclareMathDelimiter{\rclose}{\mathclose}{operators}{"5D}{largesymbols}{"03}
\DeclareMathDelimiter{\lopen}{\mathopen}{operators}{"5D}{largesymbols}{"03}
\DeclareMathDelimiter{\ropen}{\mathclose}{operators}{"5B}{largesymbols}{"02}
\newcommand{\clcl}[1]{\lclose#1\rclose}

\newcommand{\abs}[1]{\lvert#1\rvert}

\newcommand{\ketbra}[2]{\lvert#1\rangle\langle#2\rvert}

\newcommand{\expe}[1]{\langle#1\rangle}
\newcommand{\set}[1]{\{#1\}}
\newcommand{\pf}{p}
\newcommand{\sect}{\S}
\newcommand{\eqn}{eq.}%
\theoremstyle{remark}

\theoremstyle{definition}

\newcommand{\ie}{{i.e.}}

\newcommand{\eg}{{e.g.}}

\newcommand{\cf}{{cf.}}

\newcommand{\etal}{{et al.}}

\newcommand{\bd}{\hspace{0pt}}%

\providecommand{\href}[2]{#2}
\providecommand{\eprint}[2]{\texttt{\href{#1#2}{#2}}}
\renewcommand{\eprint}[2]{\texttt{\href{#1#2}{#2}}}
\newcommand{\arxiveprint}[1]{arxiv eprint
\eprint{http://arxiv.org/abs/}{#1}}

\alleugreek

\renewenvironment{acknowledgements}{\section*{Acknowledgements}}{\par}
\newcommand{\povm}{positive-\bd operator-\bd valued measure}

\newcommand{\so}{statistical operator}

\newcommand{\tsum}{{\textstyle\sum}}
\DeclareMathOperator*{\tprod}{\textstyle\prod}

\newcommand{\zrh}{\bm{\rho}}
\newcommand{\zrhha}{\hat{\bm{\rho}}}
\newcommand{\zrhh}{\rho}
\newcommand{\ze}{\bm{\varEpsilon}}

\newcommand{\zone}{\ketbra{1}{1}}

\newcommand{\zzero}{\ketbra{0}{0}}
\newcommand{\zmone}{\ketbra{-1}{-1}}

\newcommand{\zlgm}{\bm{\lambda}}
\newcommand{\zla}{x}
\newcommand{\zll}{\bm{x}}

\newcommand{\zllc}{\hat{\bm{x}}}
\newcommand{\zlac}{\hat{x}}

\newcommand{\zNv}{\bar{N}}

\newcommand{\zgi}{g_\text{co}}
\newcommand{\zgj}{g_\text{ga}}
\newcommand{\zgk}{g_\text{sl}}
\newcommand{\zI}{I_\text{co}}
\newcommand{\zJ}{I_\text{ga}}
\newcommand{\zK}{I_\text{sl}}

\newcommand{\zsoset}{\mathbb{S}}
\newcommand{\zsosett}{{\mathbb{S}_3}}
\newcommand{\zxset}{{\mathbb{B}_8}}
\newcommand{\zcset}{{\mathbb{C}_8}}

\newcommand{\zchf}{\chi_{\mathbb{B}_8}}

\newcommand{\zm}{\bar{m}}
\newcommand{\zms}{\bar{m}^*}

\newcommand{\origo}{\bm{O}}

\newcommand{\zS}{(\det{\bm{\rho}})^{2d+1}}

\newcommand{\zq}{q}
\newcommand{\zqq}{\bm{\zq}}
\newcommand{\zf}{\bm{f}}

\newcommand{\zzf}{\varPhi}
\newcommand{\zza}{\varUpsilon}
\newcommand{\zzfi}{{\varPhi_\infty}}
\newcommand{\zzfa}{{\varUpsilon_\infty}}

\begin{document}
\bibliographystyle{apsrevmananum} 

\title{Numerical Bayesian quantum-state 
       assignment for a three-level quantum system
  \\ II. Average-value data with a constant, a
  Gaussian-like, and a Slater prior}

\author{\firstname{A.} \surname{M{\aa}nsson}}
\email{andman@imit.kth.se}

\author{\firstname{P. G. L.} \surname{Porta Mana}}

\author{\firstname{G.} \surname{Bj\"{o}rk}} 

\affiliation{Kungliga Tekniska H\"ogskolan, Isafjordsgatan
  22, SE-164\,40 Stockholm, Sweden}

\date{14 January 2007}

\begin{abstract}
  This paper offers examples of concrete numerical
  applications of Bayesian quantum-state assignment
  methods to a three-level quantum system. The \so\ 
  assigned on the evidence of various measurement data 
  and kinds of prior knowledge is computed partly
  analytically, partly through numerical integration (in
  eight dimensions) on a computer. The measurement data
  consist in the average of outcome values of $N$
  identical von~Neumann projective measurements performed
  on $N$ identically prepared three-level systems. In 
  particular the large-$N$ limit will be considered. Three 
  kinds of prior knowledge are used: one represented 
  by a plausibility distribution constant in respect of
  the convex structure of the set of \so s; another one
  represented by a prior studied by Slater, which has 
  been proposed as the natural measure on the set of 
  statistical operators; the last prior is represented 
  by a Gaussian-like distribution 
  centred on a pure \so, and thus reflecting a situation 
  in which one has useful prior knowledge about the likely 
  preparation of the system. The assigned statistical 
  operators obtained with the first two kinds of priors 
  are compared with the one obtained by Jaynes' maximum 
  entropy method for the same measurement situation. 
  
  In the companion paper the case of measurement data 
  consisting in absolute frequencies is considered.
\end{abstract}


\maketitle

\section{Introduction}
\label{sec:intro}

In this paper we continue our two-part study \citep{maanssonetal2006}   
with examples of concrete numerical applications of
Bayesian quantum-state assignment methods to a three-level 
quantum system. Since we will consider the same scenario
as in the first paper, to avoid repeating ourselves we
therefore refer the reader to the first paper for a more
detailed and complete account of the motivations, explanations, 
discussions and references on the background, theory, formulas, 
nomenclature, etc, used in this paper. The main difference
between the two papers lies in the type of measurement
data considered. In the first paper the measurement data
consisted in absolute frequencies of the outcomes of $N$
identical von~Neumann projective measurements performed on
$N$ identically prepared three-level systems. Here we will
consider the same measurement situation, but the
measurement data will instead be in the form of an average of
values being associated to the measurement outcomes, in
particular $1$, $0$, and $-1$. The statistical operator encoding 
the average value data and prior knowledge is computed partly
numerically and partly analytically in the limit when $N\to\infty$, 
for a constant, and also for two different kinds of a
non-constant, prior probability distribution, and 
different average value data. A reason for studying data
of this kind, other than the obvious one that it may have
been given to us in this form, is that it constitutes  
an example of more complex data than mere absolute frequencies. 
It is also interesting to study this particular kind of 
data since it enables us to compare our assigned
statistical operators with those obtained by instead using 
Jaynes' maximum entropy method \cite{jaynes1957b} for the 
same measurement situation. The reason for doing this is
that we want to investigate whether or not this method could be 
seen as a special case of Bayesian quantum-state assignment, 
and if so, try to find the prior that would lead to the same 
\so\ as the one obtained by using the maximum entropy method.
 
\section{The present study}
\label{sec:presstud}

In this paper we study data $D$ and prior 
knowledge $I$ of the following kind:
\begin{itemize}
\item The measurement data $D$ consist in the average
  of $N$ outcome values of $N$ instances of the same measurement
  performed on $N$ identically prepared systems. The
  measurement is represented by the extreme \povm\ (\ie,
  non-degenerate `von~Neumann measurement') having three
  possible distinct outcomes $\set{\text{`1'}, \text{`2'},
    \text{`3'}}$ represented by the eigenprojectors
  $\set{\ketbra{1}{1}, \ketbra{0}{0}, \ketbra{-1}{-1}}$,
  where the eigenprojectors are labelled by their
  associated outcome values $\{1,0,-1\}$, respectively. 
  We consider the limiting case of very large $N$.
\item Three different kinds of prior knowledge $I$ are used.
  Two of them, $\zI$ and $\zJ$, are the same as those
  given in first paper, i.e. a prior plausibility
  distribution
\begin{equation}
\label{eq:first_prior_rho}
  \pf(\zrh \cond \zI)\, \di\zrh 
= \zgi(\zrh)\, \di\zrh
  \propto \di\zrh,
\end{equation} 
which is constant in respect of the convex structure of
the set of statistical operators, in the sense explained
in~\citep[\sect~3,4]{maanssonetal2006}; and a spherically 
symmetric, Gaussian-like prior distribution
\begin{equation}\label{eq:second_prior_rho}
    \pf(\zrh \cond \zJ)\, \di\zrh
= \zgj(\zrh)\, \di\zrh
\propto
    \exp\biggl\{
-\frac{\tr[(\zrh -\zrhha)^2]}{s^2}
\biggr\}\, \di\zrh,
\end{equation} 
centred on the \so\ $\zrhha$. The latter prior expresses
some kind of knowledge that leads us to assign a higher
plausibility to regions in the vicinity of $\zrhha$. 
For this prior we consider two examples, when 
$\zrhha=\ketbra{1}{1}$ and $\zrhha=\ketbra{0}{0}$.\footnote{ 
Note that the case $\zrhha=\ketbra{-1}{-1}$ is equivalent 
to the case with $\zrhha=\ketbra{1}{1}$.}

The third kind of prior knowledge, $\zK$, 
is represented by the prior plausibility distribution
\begin{equation}\label{eq:slater_prior_rho}
    \pf(\zrh \cond \zK)\, \di\zrh
= \zgk(\zrh)\, \di\zrh
\propto \zS\, \di\zrh,
\end{equation}
the so called ``Slater prior'' for a $d$-level system, which
has been proposed as a candidate for being the appropriate
measure on the set of statistical operators.~\citep{slater1995b}
\end{itemize}

The paper is organised as follows: In 
\sect~\ref{sec:scen_gen_case} we present the reasoning
leading to the statistical-operator-assignment formulae
in the case of average value data, for finite $N$ and in 
the limit when $N\to\infty$. We arrived at the same formulae 
(as special cases of formulae applicable to generic, not 
necessarily quantum-theoretical systems) in a series of
papers~\citep{portamanaetal2006,portamanaetal2006b,portamanaetal2006c}. 
In \sect~\ref{sec:exstasn3level} 
we present the particular case studied in this paper and 
give the statistical-operator-assignment formulae in this 
case, introduce the Bloch vector parametrisation, present the
calculations by symmetry arguments and by numerical integration,  
discuss the result and in some cases compare it with that  
obtained by the maximum entropy method. Finally, the last 
section summarises and discusses the main points and results.

\section{Statistical operator assignment}
\label{sec:scen_gen_case}

\subsection{General case}
\label{sec:scen_gen_case_fin}

Again we assume there is a preparation scheme that produces quantum
systems always in the same `condition' --- the same
`state' --- where each condition is associated with a
\so. Suppose we come to know that $N$ measurements,
represented by the $N$ \povm s $\set{\ze^{(k)}_\mu \with
  \mu = 1, \dotsc, r_k}$, $k= 1, \dotsc, N$, are or have
been performed on $N$ systems for which our knowledge $I$
holds. In this paper we will be analysing the case 
when the data is an average of a number of outcome 
values and it will therefore be 
natural to limit ourselves to the situation when the $N$ 
measurements are all instances of the same 
measurement. Thus, for all $k= 1, \dotsc, N$,
$\set{\ze^{(k)}_\mu} = \set{\ze_\mu}$. 

Let us say that the outcomes $i_1, \dotsc, i_k, \dotsc, i_N$ are or were
obtained. Since every outcome is associated to an outcome
value $m_i$, the average of all outcome values is 
\begin{equation}
\zm \equiv \sum_{k=1}^{N} m_{i_k}/N. 
\end{equation} 
We will consider the general situation in which the data $D$
consists in the knowledge that the average value $\zm$ in 
$N$ repetitions of the measurement lies in a set $\varUpsilon$; 
\begin{equation}
D \corr [\zm\in\varUpsilon]. 
\end{equation} 
Such kind of data arise when the 
the measurements is affected by uncertainties and is moreover
``coarse-grained'' for practical purposes, so that not
precise average values are obtained but rather a region of
possible ones. On the evidence of $D$ we can update the prior 
plausibility distribution 
$g(\zrh)\,\di\zrh\defd\pf(\zrh \cond I)\,\di\zrh$. By the rules of 
plausibility theory\footnote{We do not explicitly write the prior
  knowledge $I$ whenever the \so\ appears on the
  conditional side of the plausibility; \ie, $\pf(\cdot
  \cond \zrh) \defd \pf(\cdot \cond \zrh, I)$.}
\begin{gather}
\label{eq:update_prior}
     \pf(\zrh \cond D \land I)\,\di\zrh 
=
 \frac{ \pf(D
      \cond \zrh)\, g(\zrh)\, \di\zrh } {\int_{\zsoset}
      \pf(D \cond \zrh)\, g(\zrh)\, \di\zrh},
 \end{gather}
where $\zsoset$ is the set of all statistical operators. 

The plausibility of obtaining a particular sequence of 
outcomes is
\begin{gather}
\label{eq:many_meas2}
\pf\bigl(\ze_{i_1}, \dotsc, \ze_{i_N} \cond
   \zrh\bigr) = 
\tprod_{i=1}^r [\tr\bigl\{\ze_i \zrh \bigr\}]^{N_i},
\end{gather}
with the convention, here and in the following, that 
only factors with $N_i>0$ are to be multiplied over,
and where we have used that 
$\tr\bigl\{\ze_i \zrh \bigr\}=\pf\bigl(\ze_i \cond \zrh\bigr)$
and $(N_1,..,N_r)\defs\zNv$ are the absolute frequencies of
appearance of the $r$ possible outcomes (naturally, $N_i 
\ge 0$ and $\tsum_i N_i = N$). Since the exact order 
of the sequence of outcomes is unimportant and
only the absolute frequencies of appearance $\zNv$ matter, 
the plausibility of the absolute frequencies $\zNv$ in $N$ 
measurements is 
\begin{gather}
\label{eq:pnxfirstpaper}
\pf\bigl(\zNv \cond \zrh \bigr)
=
N! \prod_{i=1}^r \frac{[\tr\bigl\{\ze_i \zrh \bigr\}]^{N_i}}{N_i!}.
\end{gather}
Define $\mathbb{N}_N^r$ as the set of all absolute 
frequencies $\zNv$, for fixed $N$ and $r$. By the 
rules of plausibility theory we then have that
\begin{gather}
\label{eq:postinv}
\pf\bigl(D \cond \zrh \bigr)
= 
\sum_{\zNv\in\mathbb{N}_N^r} \pf\bigl(D|\zNv \land \zrh)
\pf\bigl(\zNv|\zrh).
\end{gather}
Given that we know $\zNv$, we can with certainty 
tell if $\zNv$ corresponds to an average value
\begin{equation} 
\zm \equiv \sum_{i=1}^{r} N_i m_i/N 
\end{equation}
that belongs to the set $\varUpsilon$, and knowledge of the statistical
operator $\zrh$ is here irrelevant. We thus have that 
$\pf\bigl(D|\zNv \land \zrh)=\pf\bigl(\zm\in\varUpsilon|\zNv)=1$
if $\zNv\in\phi_{\varUpsilon}$ and $0$ otherwise, where we
have defined
\begin{gather}
\phi_{\varUpsilon} 
\defd
\{\zNv\in\mathbb{N}_N^r \cond {\textstyle\sum}_i N_i
m_i/N \in \varUpsilon\}.
\end{gather}
Using this together with equations \eqref{eq:pnxfirstpaper} 
and \eqref{eq:postinv} we obtain:
\begin{gather}
\pf(D|\zrh)
=
\sum_{\zNv\in\phi_{\varUpsilon}} \pf(\zNv|\zrh)
=
N! \sum_{\zNv\in\phi_{\varUpsilon}} {\prod_{i=1}^{r}}
\frac{[\tr\{\ze_{i}\,\zrh \}]^{N_i}}{N_i!}.
\end{gather}
Inserting this into equation \eqref{eq:update_prior} 
we finally obtain: 
\begin{gather}
p(\zrh|D \land I)\, \di\zrh
= 
\frac{\displaystyle \sum_{\zNv\in\phi_{\varUpsilon}}
  \Bigl({\prod_{i=1}^{r}}
\frac{[\tr\{\ze_{i}\,\zrh\}]^{N_i}}{N_i!}\Bigr) g(\zrh)\,\di\zrh}
{\displaystyle
  \sum_{\zNv\in\phi_{\varUpsilon}}\int\limits_{\mathbb{S}}
  \Bigl({\prod_{i=1}^{r}}
\frac{[\tr\{\ze_{i}\,\zrh\}]^{N_i}}{N_i!}\Bigr) g(\zrh)\,\di\zrh}.
\end{gather}
We saw in the first paper that generic knowledge $\tilde{I}$ 
can be represented by or ``encoded in'' a unique \so: 
\begin{gather}
\zrh_{\tilde{I}}
\defd
\int_{\zsoset} \zrh\, \pf(\zrh \cond \tilde{I})\, \di\zrh.
\label{eq:def_rho_I}
\end{gather}
The \so\ encoding the joint knowledge $D \land I$ is thus
given by
\begin{gather}
\label{eq:genassso}
\bm{\rho}_{D \land I} 
=
\frac{\displaystyle\sum_{\zNv\in\phi_{\varUpsilon}}\int\limits_{\mathbb{S}} \zrh\,\Bigl({\prod_{i=1}^{r}}
\frac{[\tr\{\ze_{i}\zrh\}]^{N_i}}{N_i!}\Bigr) g(\zrh)\,\di\zrh}
{\displaystyle\sum_{\zNv\in\phi_{\varUpsilon}}\int\limits_{\mathbb{S}} \Bigl({\prod_{i=1}^{r}}
\frac{[\tr\{\ze_{i}\zrh\}]^{N_i}}{N_i!}\Bigr) g(\zrh)\,\di\zrh}.
\end{gather}

\subsection{Large-$N$ limit}
\label{sec:scen_gen_case_infty}

Let us now summarise some results obtained 
in~\citep{portamanaetal2006c} for the case of very large $N$. 
Consider the general situation in which each data set $D_N$ 
consists in the knowledge that the relative frequencies 
$\zf\equiv(f_i):=(N_i/N)$
lie in a region $\zzf_N=\{\zf\cond[\sum_if_i\,m_i]\in\zza_N\}$, 
where $\zza_N$ is a region in which the average values
lie (being such that $\zzf_N$ has a non-empty interior and its
boundary has measure zero in respect of the prior
plausibility measure). Mathematically we want to
see what form the state-assignment formulae take in the
limit $N \to \infty$. Consider a sequence of data sets 
$\set{D_N}_{N=1}^\infty$ with corresponding sequences of 
regions $\set{\zza_N}_{N=1}^\infty$ and $\set{\zzf_N}_{N=1}^\infty$,
and assume the regions converges (in a topological sense specified 
in~\citep{portamanaetal2006c}) to regions $\zzfa$ and 
$\zzfi$ (the latter also with non-empty interior and with
boundary of measure zero), respectively. 

Given that the \so\ 
is $\zrh$, the plausibility distribution for the outcomes is
\begin{equation}
  \label{eq:plaus_out_N}
  \zqq(\zrh)
  \equiv \bigl(\zq_i(\zrh)\bigr)\quad \text{with} \quad \zq_i(\zrh) \defd \tr(\ze_i \zrh).
\end{equation}
In \citep{portamanaetal2006c} it is shown that
\begin{multline}
  \label{eq:post_N}
  \pf(\zrh \cond  D_N \land I)\, \di\zrh
\propto 
\begin{cases}
  0,& \text{if $\zqq(\zrh) \not\in \zzfi$},
\\
 \pf(\zrh \cond I)\, \di\zrh,
& \text{if  $\zqq(\zrh) \in \zzfi$},
\end{cases}
\\ \text{as $N \to \infty$}.
\end{multline}
Further it is also shown that if $\zzfa$ degenerates into
a single average value $\zm$, the expression above
becomes\footnote{Note that we have here, with abuse of notation, 
  written $\pf[\zrh \cond  \zm \land I]$ instead of the more 
  correct form $\pf[\zrh \cond (\zm=\zms) \land I]$, to avoid 
  introducing another variable $\zms$ for the average
  value data.}
\begin{equation}
  \label{eq:post_N_delt}
  \pf[\zrh \cond  \zm \land I]\, \di\zrh
\propto
\pf(\zrh \cond I)\, \delt[{\textstyle\sum}_i\zq_i(\zrh)\,m_i-\zm]\,\di\zrh.
\end{equation}
This is an intuitively satisfying result, since in the limit 
when $N\to\infty$ we would expect that it is only 
those statistical operators $\zrh$ whose expectation value 
${\textstyle\sum}_i\zq_i(\zrh)\,m_i$ is equal to the
measured average value $\zm$ that could have been
the case. The data single out a set of statistical
operators, and these are then given weight according to
the prior $\pf(\zrh\cond I)\,\di\zrh = g(\zrh)\,\di\zrh$, 
specified by us. 

By normalising the posterior plausibility 
distribution in equation~\eqref{eq:post_N_delt}, the assigned statistical 
operator in equation~\eqref{eq:def_rho_I} is then given by
\begin{gather}
\label{eq:ass_op_ninfty}
\bm{\rho}_{D \land I} 
=
\frac{\displaystyle \int\limits_{\mathbb{S}} 
\zrh\,g(\zrh)\,\delt[{\textstyle\sum}_i\zq_i(\zrh)\,m_i-\zm]\,\di\zrh}
{\displaystyle \int\limits_{\mathbb{S}} 
g(\zrh)\,\delt[{\textstyle\sum}_i\zq_i(\zrh)\,m_i-\zm]\,\di\zrh}.
\end{gather}

\section{An example of state assignment for a three-level system}
\label{sec:exstasn3level}

\subsection{Three-level case}
\label{sec:exstasn3level_3levcase}

We will now consider the particular case studied in this paper.
The preparation scheme concerns three-level quantum
systems; the corresponding set of \so s will be denoted by
$\zsosett$.  We are going to consider the case when the number 
of measurements $N$ is very large and in the limit 
goes to infinity. The $N$ measurements are all instances of the same 
measurement, namely a non-degenerate projection-valued 
measurement (often called `von~Neumann measurement'). 
Thus, for all $k= 1, \dotsc, N$,
$\set{\ze^{(k)}_\mu} = \set{\ze_\mu} \defd \set{\zone,
\zzero, \zmone}$, where the projectors, labelled by the particular outcome
values $(m_1,m_2,m_3)=(1,0,-1)$ we have chosen to consider here, 
define an orthonormal basis in Hilbert space. All relevant
operators will, quite naturally and advantageously, be
expressed in this basis. We have for example that
$\zq_{\mu}(\zrh)=\tr(\ze_{\mu} \zrh) = \zrhh_{\mu \mu}$, the $\mu$th
diagonal element of $\zrh$ in the chosen basis. As data we 
are given that the average of the
measurement outcome values is $\zm$ (more precisely in
the sense that $\zzfi$ degenerates into a single average
value $\zm$).

\subsection{Bloch vector parametrisation and symmetries}
\label{sec:exstasn3level_blvecsym}

We will be using the same parametrisation 
of the statistical operators as in 
the companion paper, i.e. in terms of Bloch vectors
$\zll$. For a three-level system the Bloch vector expansion 
of a \so\ $\zrh(\zll)$ is given by:
\begin{equation}
\label{eq:blvexp}
\zrh(\zll) = 
 \frac{1}{3} \bm{I}_3 + 
 \frac{1}{2}\textstyle{\sum}_{j=1}^{8} \zla_j \zlgm_j,
\end{equation} 
where
\begin{equation}
\label{eq:blvcoef}
\zla_i
=
\tr\{\zlgm_i\,\zrh(\zll)\}
\equiv
\expe{\zlgm_i}_{\zrh(\zll)}.
\end{equation} 
The Gell-Mann operators $\zlgm_i$ are Hermitian and can 
therefore be regarded as observables. Note that 
our von~Neumann measurement corresponds to 
the observable 
\begin{equation}
\zlgm_3\equiv\zone + 0\, \zzero - \zmone.\label{eq:lambda_3-explic}
\end{equation}
Hence, given a \so\ $\zrh(\zll)$, the following holds 
for the expectation value of the outcome values 
$\{1,0,-1\}$ for this particular measurement:
\begin{gather}
\expe{\zlgm_i}_{\zrh(\zll)}
=
{\textstyle\sum}_i\zq_i(\zll)\,m_i 
=
\zrhh_{11}(\zll)-\zrhh_{33}(\zll)
=
\zla_3.
\end{gather}
Equation~\eqref{eq:post_N_delt} thus becomes
\begin{equation}
  \label{eq:post_N_delt_bvp}
  \pf[\zll \cond \zm \land I]\, \di\zll
\propto
g(\zll)\, \delt(\zla_3-\zm)\,\di\zll,
\end{equation}
and the assigned statistical operator in 
equation~\eqref{eq:ass_op_ninfty} assumes the form
 \begin{gather}
\label{eq:ass_op_ninfty_b}
\bm{\rho}_{\zm \land I} 
=
\frac{\displaystyle \int\limits_{\zxset} 
\zrh(\zll)\,g(\zll)\,\delt(\zla_3-\zm)\,\di\zll}
{\displaystyle \int\limits_{\zxset} 
g(\zll)\,\delt(\zla_3-\zm)\,\di\zll},
\end{gather}
where $\zxset$ is the set of all three-level Bloch vectors.
This can be rewritten in a form especially suited for numerical 
integration by computer, which we shall use hereafter:
\begin{gather}
\label{eq:statopass}
\bm{\rho}_{\zm \land I} 
= 
\frac{\displaystyle\int\limits_{\zcset} 
\zrh(\zll) g(\zll)\delt(\zla_3-\zm)\,\zchf(\zll)\,\di\zll}
{\displaystyle\int\limits_{\zcset} 
g(\zll)\delt(\zla_3-\zm)\,\zchf(\zll)\,\di\zll},
\end{gather}
where $\zchf(\zll)$ is the characteristic
function of the set $\zxset\subset\zcset \defd \clcl{-1,1}^7 \times
 \Bigl\lclose-\tfrac{2}{\sqrt{3}}, \tfrac{1}{\sqrt{3}}
 \Bigr\rclose$. Using the Bloch vector expansion in 
equation~\eqref{eq:blvexp} we see that by computing the
following set of integrals we have determined 
$\bm{\rho}_{\zm \land I}$:
\begin{equation}
\label{eq:integrals}
L_i[\zm,I] \defd \int\limits_{\zcset} 
\zla_i\, g(\zll)\, \delt(\zla_3-\zm)\,\zchf(\zll)\,\di\zll,
\end{equation}   
where $i\in\{1,..,8\}$, and
\begin{equation}
\label{eq:integralsz}
Z[\zm,I] \defd \int\limits_{\zcset} 
g(\zll)\, \delt(\zla_3-\zm)\,\zchf(\zll)\,\di\zll,
\end{equation}
where the dependence of the average value and prior knowledge 
is indicated within brackets.
The assigned statistical operator will then given by
\begin{equation}
\label{eq:genso}
\bm{\rho}_{\zm \land I}  \,=\, \frac{1}{3} \bm{I}_3 + \frac{1}{2}
\sum_{i=1}^{8} \frac{L_i[\zm,I]}{Z[\zm,I]} \bm{\lambda}_i. 
\end{equation} 
One sees directly from equations \eqref{eq:integrals} and 
\eqref{eq:integralsz} that $L_3[\zm,I]/Z[\zm,I] = \zm$ 
($Z[\zm,I]$ can never vanish, its integrand being positive
and never identically naught). 

For the same reasons already accounted for in the first paper we will not
try to determine $\bm{\rho}_{\zm \land I}$ exactly, but
also here compute it with a combination of symmetry 
considerations of $\zxset$ and numerical integration. 
For all three kinds of prior knowledge 
considered in this paper the same symmetry arguments used
in the companion paper also holds here, so again we have 
that $L_i[\zm,I]/Z[\zm,I]=0$ for all $i\neq3,8$ and any
average value $-1 \leq \zm \leq 1$. The assigned Bloch vector is thus
given by $(0,0,\zm,0,0,0,0,L_8[\zm,I]/Z[\zm,I])$. 
This means that $\zrh_{\zm \land I}$ lies in the 
$(\zla_3,\zla_8)$-plane and it has, in the chosen 
eigenbasis, the diagonal matrix form
\begin{equation}
\label{eq:rhomatrix_diag}
\zrh_{\zm \land I}=
\begin{pmatrix}
                      \frac{1}{3}+\frac{\zm}{2}+ \frac{L_8[\zm,I]}{2\sqrt{3} Z[\zm,I]}
                    & 0
                    &0 \\
                     0  
                    & \frac{1}{3}-\frac{L_8[\zm,I]}{\sqrt{3} Z[\zm,I]}
                    & 0 \\
                     0
                    & 0
                   & \frac{1}{3}-\frac{\zm}{2}+
                   \frac{L_8[\zm,I]}{2\sqrt{3} Z[\zm,I]}
               \end{pmatrix}.
\end{equation} 

\subsection{Numerical integration, results and the maximum entropy method}
\label{sec:exstasn3level_numintresme}

\begin{figure*}[p]
\includegraphics[width=2\columnwidth]{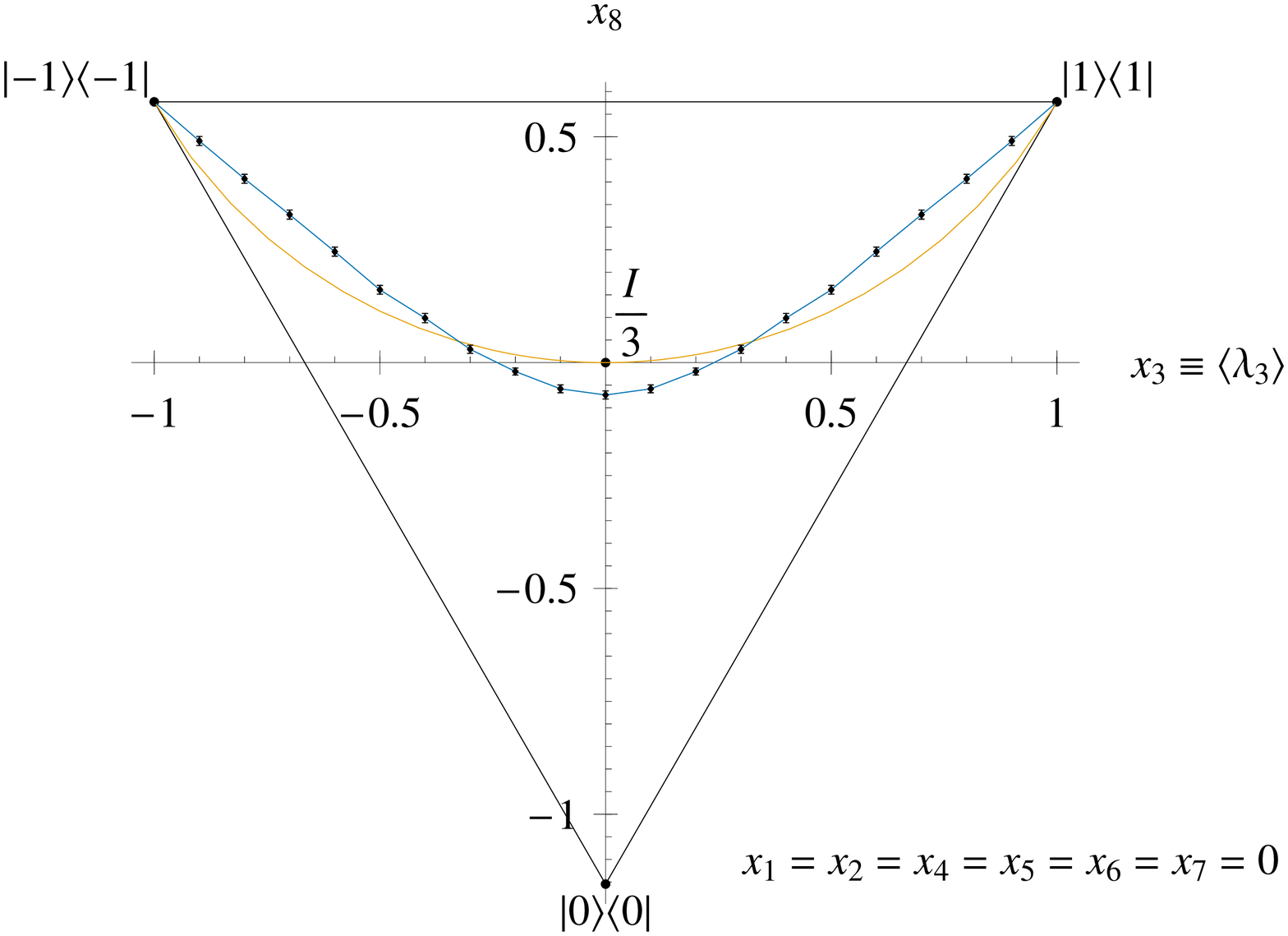}
\caption{Bloch vectors of the assigned \so\ for prior 
  knowledge $\zI$, computed by numerical integration 
  for different average values 
  $\zm\equiv\zla_3\equiv\expe{\zlgm_3}$
  (connected by the blue curve [with bars]). The red curve 
  (without bars) is the statistical operator given by the maximum 
  entropy method also as a function of the average value 
  $\zm\equiv\zla_3\equiv\expe{\zlgm_3}$. The large triangle 
  is the two-dimensional cross section of the set 
  $\zxset$ along the plane $\origo\zla_3\zla_8$. The 
  maximum numerical-integration uncertainty in the $\zla_8$ 
  component is $\pm 0.01$. Note that only the ten points 
  for $0\leq\zm\leq0.9$ have been determined by numerical integration,
  since the nine points for $-0.9\leq\zm\leq-0.1$ can be exactly 
  determined from the former by symmetry arguments. The 
  endpoints corresponding to $\zm=\pm 1$ were set manually, 
  since $\zla_8=1/\sqrt{3}$ is the only possibility in this 
  case. Within the given uncertainties, numerical 
  computations yielded the exact results.
  The computation was done on a PC (Pentium $4$ processor,
  $3$~GHz) and the computation time was $15$~min.
\label{fig:Fig_const}}
\end{figure*}

\begin{figure*}[p]
\includegraphics[width=2\columnwidth]{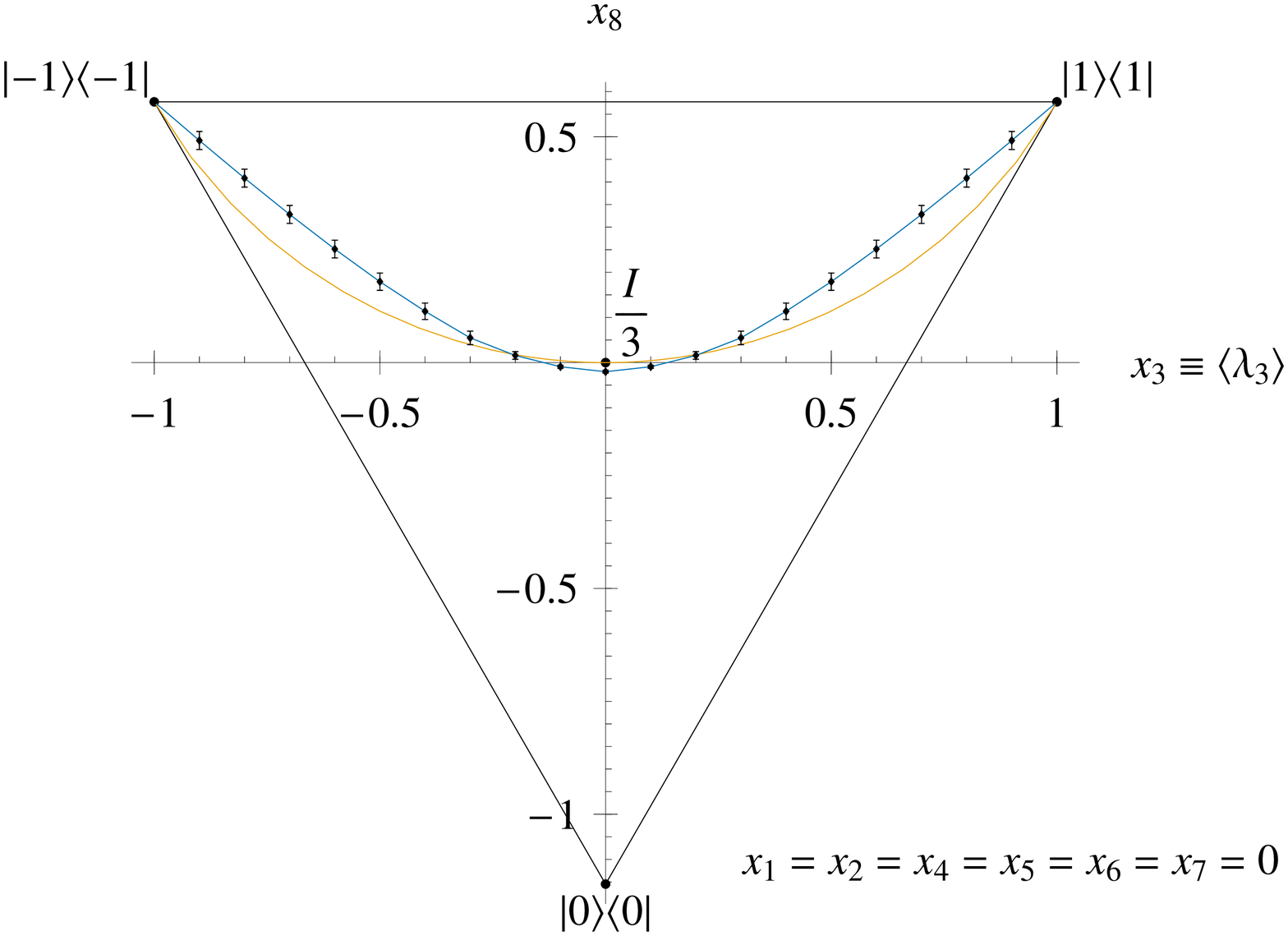}
\caption{Bloch vectors of the assigned \so\ for prior 
  knowledge $\zK$, computed by numerical integration 
  for different average values 
  $\zm\equiv\zla_3\equiv\expe{\zlgm_3}$
  (connected by the blue curve [with bars]). The red curve 
  (without bars) is the statistical operator given by the maximum 
  entropy method also as a function of the average value 
  $\zm\equiv\zla_3\equiv\expe{\zlgm_3}$. The large triangle 
  is the two-dimensional cross section of the set 
  $\zxset$ along the plane $\origo\zla_3\zla_8$. The 
  maximum numerical-integration uncertainty in the $\zla_8$ 
  component is $\pm 0.02$. Note that only the ten points 
  for $0\leq\zm\leq0.9$ have been determined by numerical integration,
  since the nine points for $-0.9\leq\zm\leq-0.1$ can be exactly 
  determined from the former by symmetry arguments. The 
  endpoints corresponding to $\zm=\pm 1$ were set manually, 
  since $\zla_8=1/\sqrt{3}$ is the only possibility in this 
  case. Within the given uncertainties, numerical 
  computations yielded the exact results. The computation
  was done on a PC (Pentium $4$ processor,
  $3$~GHz) and the computation time was $250$~min.
\label{fig:Fig_slater}}
\end{figure*}

\begin{figure*}[p]
\includegraphics[width=2\columnwidth]{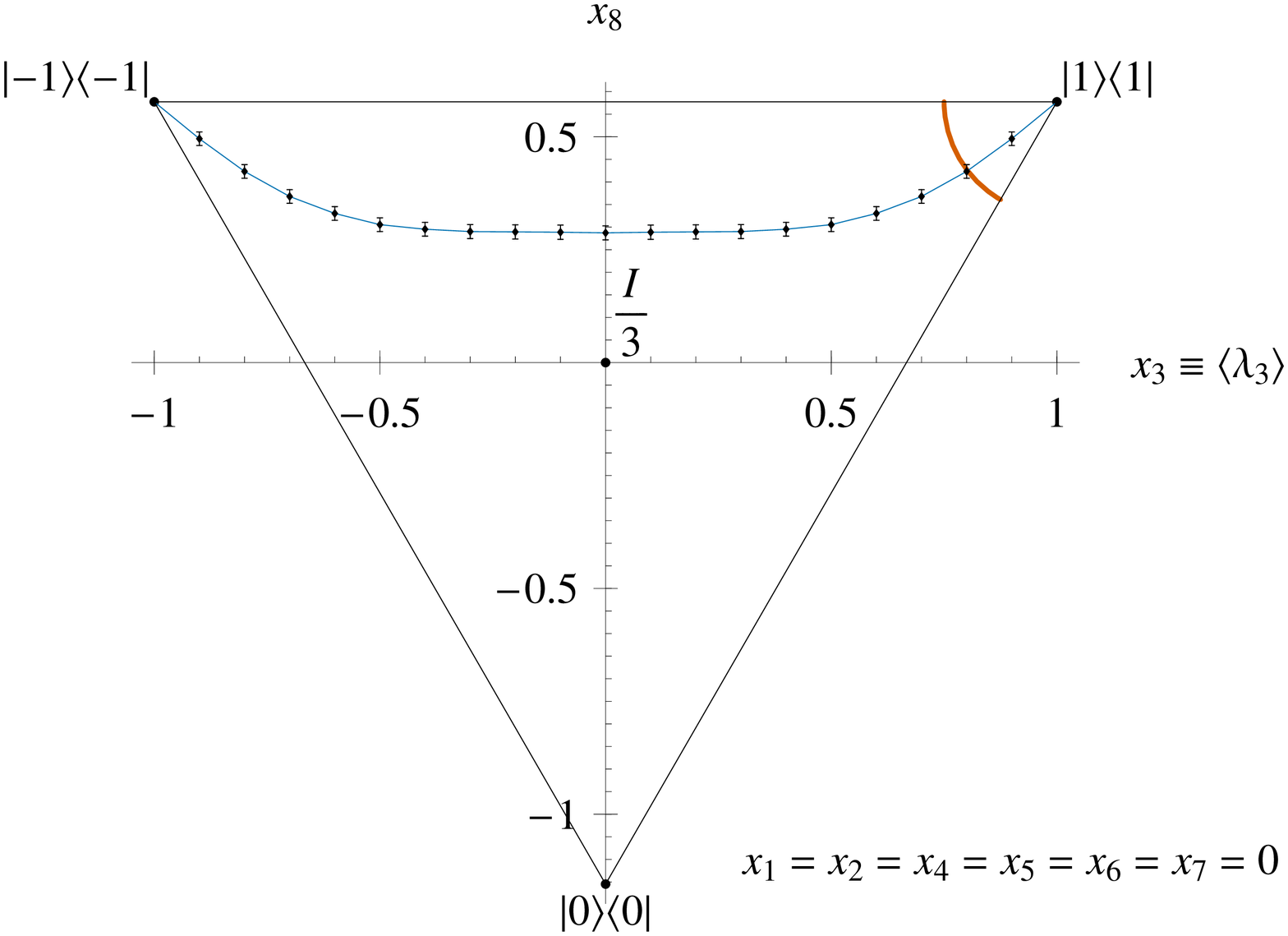}
\caption{Bloch vectors of the assigned \so\ for prior 
  knowledge $\zJ$, computed by numerical integration 
  for different average values 
  $\zm\equiv\zla_3\equiv\expe{\zlgm_3}$ (connected 
  by the curve). The large triangle 
  is the two-dimensional cross section of the set 
  $\zxset$ along the plane $\origo\zla_3\zla_8$. 
  The prior knowledge is represented by a Gaussian-like 
  distribution of ``breadth'' $s=1/4$ centred 
  on the pure \so\ $\zone$. The small circular arc is the
  locus of the Bloch vectors (on the plane) at a distance
  $\abs{\zll - \zllc} =s$ from the vector $\zllc \defd
  (0,0,1,0,0,0,0,1/\sqrt{3})$ corresponding to the \so\
  $\zone$. The numerical-integration uncertainty in the 
  $\zla_8$ component is $\pm 0.016$. Note that only the ten points 
  for $0\leq\zm\leq0.9$ have been determined by numerical integration,
  since the nine points for $-0.9\leq\zm\leq-0.1$ can be exactly 
  determined from the former by symmetry arguments. The
  endpoints corresponding to $\zm=\pm 1$ 
  were set manually, since $\zla_8=1/\sqrt{3}$ is the 
  only possibility in this case. Within the given 
  uncertainties, numerical computations yielded the 
  exact results. The computation
  was done on a PC (Pentium $4$ processor,
  $3$~GHz) and the computation time was $30$~min.
\label{fig:Fig_gauss1}}
\end{figure*}

\begin{figure*}[p]
\includegraphics[width=2\columnwidth]{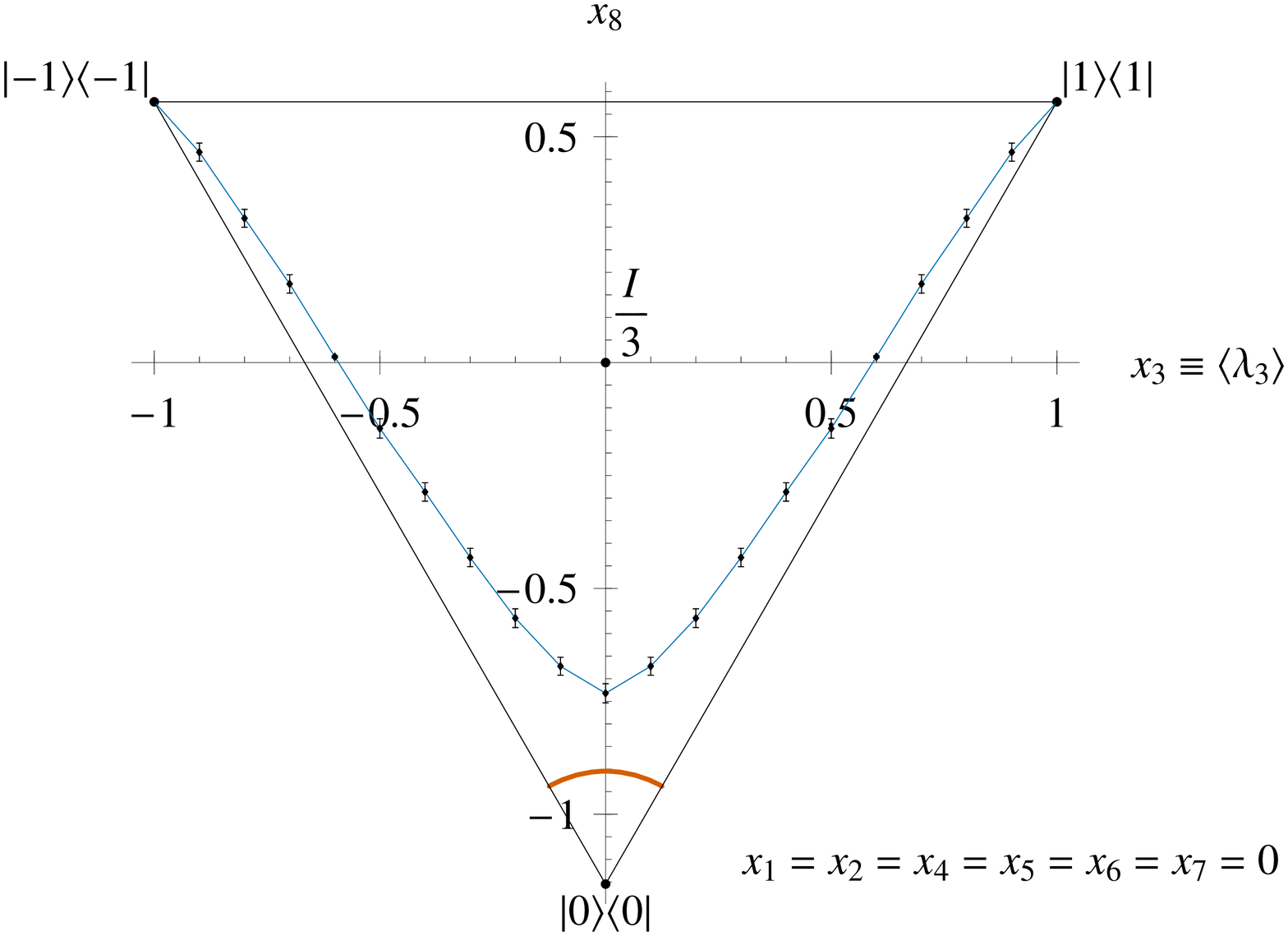}
\caption{Bloch vectors of the assigned \so\ for prior 
  knowledge $\zJ$, computed by numerical integration 
  for different average values 
  $\zm\equiv\zla_3\equiv\expe{\zlgm_3}$ (connected 
  by the curve). The large triangle 
  is the two-dimensional cross section of the set 
  $\zxset$ along the plane $\origo\zla_3\zla_8$. 
  The prior knowledge is represented by a Gaussian-like 
  distribution of ``breadth'' $s=1/4$ centred 
  on the pure \so\ $\zzero$. The small circular arc is the
  locus of the Bloch vectors (on the plane) at a distance
  $\abs{\zll - \zllc} =s$ from the vector $\zllc \defd
  (0,0,0,0,0,0,0,-2/\sqrt{3})$ corresponding to the \so\
  $\zzero$. The numerical-integration uncertainty in the 
  $\zla_8$ component is $\pm 0.02$. Note that only the ten points 
  for $0\leq\zm\leq0.9$ have been determined by numerical integration,
  since the nine points for $-0.9\leq\zm\leq-0.1$ can be exactly 
  determined from the former by symmetry arguments. The
  endpoints corresponding to $\zm=\pm 1$ 
  were set manually, since $\zla_8=1/\sqrt{3}$ is the 
  only possibility in this case. Within the given 
  uncertainties, numerical computations yielded the 
  exact results. The computation
  was done on a PC (Pentium $4$ processor, $3$~GHz) 
  and the computation time was $425$~min.
\label{fig:Fig_gauss0}}
\end{figure*}

We have used numerical integration\footnote{Using 
quasi Monte Carlo-integration in Mathematica 5.2 on a PC 
(Pentium~$4$ processor, $3$~GHz). The computation times are
given in figures~\ref{fig:Fig_const} to~\ref{fig:Fig_gauss0},
and for more details on the numerical integration we again refer the 
reader to the companion paper~\citep{maanssonetal2006}.} to 
compute $L_8[\zm,I]/Z[\zm,I]$ for different prior knowledge and different 
values of $\zm$. The result for a constant prior density is shown in
figure~\ref{fig:Fig_const}, where the blue
curve (with bars indicating the numerical-integration
uncertainties) is the Bloch vector corresponding to $\zrh_{\zm \land \zI}$ 
plotted for different values of $\zla_3=\zm$.\footnote{\label{ftn:symm}
Note that we have for all three kinds of priors considered
in this paper computed $L_8[\zm,I]/Z[\zm,I]$ only for 
non-negative values of $\zla_3=\zm$, since by using the symmetry operation 
$(\zla_1,\zla_2,\zla_3,\zla_4,\zla_5,\zla_6,\zla_7,\zla_8)
\mapsto (\zla_6,\zla_7,-\zla_3,\zla_4,-\zla_5,\zla_1,\zla_2,\zla_8)$ 
one can show that $L_8[\zm,I]/Z[\zm,I]$ is invariant under a sign
change of $\zm$. Further, we have not computed $L_8[\zm,I]/Z[\zm,I]$
for $\zm=\pm 1$, since it follows from \citep[\eqn~17]{maanssonetal2006} 
that $L_8[\zm,I]/Z[\zm,I]=1/\sqrt{3}$ is the only
possibility in this case (which one also realises by
looking at the figures).}
 
It is interesting to compare $\zrh_{\zm \land \zI}$ with
the statistical operator obtained by 
the maximum entropy method \cite{jaynes1957b} for the 
measurement situation we are considering here.
Given the expectation value $\expe{\bm{M}}$ of a 
Hermitian operator $\bm{M}$, corresponding to an 
observable $M$, the maximum entropy method assigns the 
statistical operator to the system that maximises the 
von Neumann entropy $S\defd-\tr\{\bm{\rho}\ln{\bm{\rho}}\}$ and
satisfies the constraint $\tr\{\bm{\rho}\bm{M}\}=\expe{\bm{M}}$. Having
obtained an average value $\bar{M}$ from many instances of
the same measurement performed on identically prepared
systems, one conventionally sets $\expe{\bm{M}}=\bar{M}$.
 
In our case the operator $\bm{M}$ would be 
identified as the Hermitian operator $\bm{\lambda}_3$ and
$\bar{M}$ as $\zm$. Hence the maximum entropy method
corresponds here to an assignment of the statistical
operator that maximises $S$ among all statistical operators
satisfying $\expe{\bm{\lambda}_3}=\zm$, and this 
statistical operator is given by 
\begin{gather}
\bm{\rho}_{ME} \defd \frac{\displaystyle \e^{-\mu(\zm)\,
    \bm{\lambda}_3}}{\displaystyle \tr\{\e^{-\mu(\zm) \,\bm{\lambda}_3}\}},
\end{gather}
where
\begin{gather}
\mu(\zm) := 
\ln\Biggl\{\frac{\displaystyle -\zm+\sqrt{4-3\zm^2}}{\displaystyle
  2(\zm+1)}\Biggr\}.
\end{gather}
This could be compared with the statistical operator 
$\zrh_{\zm \land I}$ obtained by instead using Bayesian 
quantum-state assignment, and expressed in general form 
as in equation~\eqref{eq:ass_op_ninfty_b} it is seen 
to instead be given by a weighted sum, with weight 
$g(\zll)\,\di\zll$, of all statistical operators with 
$\expe{\bm{\lambda}_3}=\zla_3=\zm$. 

In the case of a constant prior one sees from 
figure~\ref{fig:Fig_const} that $\zrh_{\zm \land \zI}$ 
is in general different from $\bm{\rho}_{ME}$ (the red 
curve [without bars]). This means for instance that, 
if the maximum entropy method is a special case of
Bayesian quantum-state assignment, the statistical operator obtained by the 
former method corresponds to a non-constant prior probability 
distribution $g(\zll)\,\di\zll$ on $\zxset$ in the 
latter method. This conclusion in itself is perhaps 
not so surprising, but it raises an interesting 
question: Does there exist a (non-constant on $\zxset$)
prior distribution $g(\zrh)\,\di\zrh$ that one with Bayesian
quantum-state assignment in general obtains the same assigned
statistical operator as with the maximum entropy method? 

A strong candidate is the ``Bures prior'' which has 
been proposed as the natural measure on the set of all
statistical operators (see 
e.g.~\citep{byrdetal2001,slater2001,slater2001b,slater1999b,slater1996c}), 
but unfortunately it turns out to be difficult to do numerical integrations 
on it due to its complicated functional form, so we have
not computed the assigned statistical operator in this case. 
Another interesting candidate is the ``Slater prior''~\citep{slater1995b}, 
which have also been suggested to be the natural measure on the set of all
statistical operators, and the computed assigned
statistical operator in this case is shown in 
figure~\ref{fig:Fig_slater}. One can see directly from the figure
that although it is similar to the curve obtained by the
maximum entropy method, we have found them to differ.\\ 
The computed assigned statistical operators for
the Gaussian-like prior, centred on the projectors 
$\zrhha=\ketbra{1}{1}$ and $\zrhha=\ketbra{0}{0}$ with
``breadth'' $s=1/4$, are shown in figures
\ref{fig:Fig_gauss1} and \ref{fig:Fig_gauss0},
respectively. Apart from being symmetric under a sign
change of $\zm$, as already have been noted in
footnote~\ref{ftn:symm}, one can also show that 
$L_8[\zm,\zJ]/Z[\zm,\zJ]$ does not depend on 
the $\zlac_3$-coordinate of the statistical operator 
the prior is centred on.

\section{Conclusions}
\label{sec:conclusions}

This was the second paper in a two-part study 
where the Bayesian quantum-state assignment methods has
been applied to a three-level system, showing that
the numerical implementation is possible and simple in
principle. This paper should not only be of
theoretical interest but also be of use to experimentalists 
involved in state estimation. We have
analysed the situation where we are given the average of
outcome values from $N$ repetitions of identical
von~Neumann projective measurements performed on $N$ 
identically prepared three-level systems, when 
the number of repetitions $N$ becomes very large. From this measurement 
data together with different kinds of prior knowledge of the 
preparation, a statistical operator can be assigned to the system. 
By a combination of symmetry arguments and numerical
integration we computed the assigned statistical operator 
for different average values and for a constant, 
and also for two examples of a non-constant, prior probability 
distribution.\\
The results were also compared with that obtained by 
the maximum entropy method. An interesting question is whether 
there exists a prior probability distribution that gives rise to an
assigned statistical operator which is in general
identical to the one given by the maximum entropy method,
i.e. if the maximum entropy method could be seen as a
special case of Bayesian quantum-state assignment? In the 
case of a constant and a ``Slater prior'' on the Bloch 
vector space of a three-level system 
we saw that the assigned statistical 
operator did not agree with the one given by the maximum
entropy method. It would therefore be interesting to try other 
kinds of priors, in particular
``special'' priors like the Bures one.\\ 
The generalisation of the present study to data involving
different kinds of measurement is straightforward. Of
course, in the general case one has to numerically
determine a greater number of parameters (the $L_j[\zm,I]$) and
therefore compute a greater number of integrals.
\smallskip
\paragraph{Post scriptum:}
During the preparation of this manuscript, P. Slater kindly 
informed us that some of the integrals numerically 
computed here and in the previous paper can in fact be 
calculated analytically, using \emph{cylindrical algebraic 
decomposition}~\citep{arnonetal1984,davenportetal1987_t1993,
mishra1993,jirstrand1995,brown2001b} with a parametrisation 
introduced by Bloore~\citep{bloore1976}; 
\cf~Slater~\citep{slater2006c}. This is true, \eg, for the 
integrals involving the constant and Slater's priors. By this 
method Slater has also proven the exact validity of \eqn~(52) 
of our previous paper~\citep{maanssonetal2006}. We plan to 
use and discuss more extensively this method in later 
versions of these papers. 

\begin{acknowledgements}
We cordially thank P. Slater for introducing us to cylindrical 
algebraic decomposition and showing how it can be applied to 
the integrals considered in our papers. AM thanks Professor 
Anders Karlsson for encouragement. PM thanks Louise for 
continuous and invaluable support, and the staff of the KTH 
Biblioteket for their irreplaceable work.
\end{acknowledgements}
 
(Note: `arxiv eprints' are located at
\url{http://arxiv.org/}.)

\begin{thebibliography}{19}
\expandafter\ifx\csname natexlab\endcsname\relax\def\natexlab#1{#1}\fi
\expandafter\ifx\csname bibnamefont\endcsname\relax
  \def\bibnamefont#1{#1}\fi
\expandafter\ifx\csname bibfnamefont\endcsname\relax
  \def\bibfnamefont#1{#1}\fi
\expandafter\ifx\csname citenamefont\endcsname\relax
  \def\citenamefont#1{#1}\fi
\expandafter\ifx\csname url\endcsname\relax
  \def\url#1{\texttt{#1}}\fi
\expandafter\ifx\csname urlprefix\endcsname\relax\def\urlprefix{URL }\fi
\providecommand{\bibinfo}[2]{#2}
\providecommand{\eprint}[2][]{\url{#2}}

\bibitem[{\citenamefont{M{\aa}nsson \etal}(2006)\citenamefont{M{\aa}nsson,
  Porta~Mana, and Bj\"ork}}]{maanssonetal2006}
\bibinfo{author}{\bibfnamefont{A.}~\bibnamefont{M{\aa}nsson}},
  \bibinfo{author}{\bibfnamefont{P.~G.~L.} \bibnamefont{Porta~Mana}},
  \bibnamefont{and} \bibinfo{author}{\bibfnamefont{G.}~\bibnamefont{Bj\"ork}},
  \emph{\bibinfo{title}{Numerical {Bayesian} quantum-state assignment for a
  three-level quantum system. {I}. {Absolute}-frequency
  data with a constant and a
  {Gaussian}-like prior}} (\bibinfo{year}{2006}),
  \bibinfo{note}{\arxiveprint{quant-ph/0612105}}.

\bibitem[{\citenamefont{Jaynes}(1957{\natexlab{a}})}]{jaynes1957b}
\bibinfo{author}{\bibfnamefont{E.~T.} \bibnamefont{Jaynes}},
  \emph{\bibinfo{title}{Information theory and statistical mechanics. {II}}},
  \bibinfo{journal}{Phys.\ Rev.}
  \textbf{\bibinfo{volume}{108}}(\bibinfo{number}{2}),
  \bibinfo{pages}{171--190} (\bibinfo{year}{1957}{\natexlab{a}}),
  \bibinfo{note}{\url{http://bayes.wustl.edu/etj/node1.html}, see
  also}\\
  \emph{\bibinfo{title}{Information theory and statistical mechanics}},
  \bibinfo{journal}{Phys.\ Rev.}
  \textbf{\bibinfo{volume}{106}}(\bibinfo{number}{4}),
  \bibinfo{pages}{620--630} (\bibinfo{year}{1957}{\natexlab{b}}),
  \bibinfo{note}{\url{http://bayes.wustl.edu/etj/node1.html}}.

\bibitem[{\citenamefont{Slater}(1995)}]{slater1995b}
\bibinfo{author}{\bibfnamefont{P.~B.} \bibnamefont{Slater}},
  \emph{\bibinfo{title}{Reformulation for arbitrary mixed states of {Jones}'
  {Bayes} estimation of pure states}}, \bibinfo{journal}{Physica A}
  \textbf{\bibinfo{volume}{214}}(\bibinfo{number}{4}),
  \bibinfo{pages}{584--604} (\bibinfo{year}{1995}).

\bibitem[{\citenamefont{Porta~Mana
  \etal}(2006{\natexlab{a}})\citenamefont{Porta~Mana, M{\aa}nsson, and
  Bj\"ork}}]{portamanaetal2006}
\bibinfo{author}{\bibfnamefont{P.~G.~L.} \bibnamefont{Porta~Mana}},
  \bibinfo{author}{\bibfnamefont{A.}~\bibnamefont{M{\aa}nsson}},
  \bibnamefont{and} \bibinfo{author}{\bibfnamefont{G.}~\bibnamefont{Bj\"ork}},
  \emph{\bibinfo{title}{From ``plausibilities of plausibilities'' to
  state-assignment methods: {I}. ``{Plausibilities} of plausibilities'': an
  approach through circumstances}} (\bibinfo{year}{2006}{\natexlab{a}}),
  \bibinfo{note}{\arxiveprint{quant-ph/0607111}}.

\bibitem[{\citenamefont{Porta~Mana
  \etal}(2006{\natexlab{b}})\citenamefont{Porta~Mana, M{\aa}nsson, and
  Bj\"ork}}]{portamanaetal2006b}
\bibinfo{author}{\bibfnamefont{P.~G.~L.} \bibnamefont{Porta~Mana}},
  \bibinfo{author}{\bibfnamefont{A.}~\bibnamefont{M{\aa}nsson}},
  \bibnamefont{and} \bibinfo{author}{\bibfnamefont{G.}~\bibnamefont{Bj\"ork}},
  \emph{\bibinfo{title}{From ``plausibilities of plausibilities'' to
  state-assignment methods: {II}. {Induction} and a challenge to {de~Finetti}'s
  theorem}} (\bibinfo{year}{2006}{\natexlab{b}}), \bibinfo{note}{in
  preparation}.

\bibitem[{\citenamefont{Porta~Mana
  \etal}(2006{\natexlab{c}})\citenamefont{Porta~Mana, M{\aa}nsson, and
  Bj\"ork}}]{portamanaetal2006c}
\bibinfo{author}{\bibfnamefont{P.~G.~L.} \bibnamefont{Porta~Mana}},
  \bibinfo{author}{\bibfnamefont{A.}~\bibnamefont{M{\aa}nsson}},
  \bibnamefont{and} \bibinfo{author}{\bibfnamefont{G.}~\bibnamefont{Bj\"ork}},
  \emph{\bibinfo{title}{From ``plausibilities of plausibilities'' to
  state-assignment methods: {III}. {Interpretation} of ``state'' and
  state-assignment methods}} (\bibinfo{year}{2006}{\natexlab{c}}),
  \bibinfo{note}{in preparation}.

\bibitem[{\citenamefont{Byrd and Slater}(2001)}]{byrdetal2001}
\bibinfo{author}{\bibfnamefont{M.~S.} \bibnamefont{Byrd}} \bibnamefont{and}
  \bibinfo{author}{\bibfnamefont{P.~B.} \bibnamefont{Slater}},
  \emph{\bibinfo{title}{{Bures} measures over the spaces of two- and
  three-dimensional density matrices}}, \bibinfo{journal}{Phys.\ Lett.\ A}
  \textbf{\bibinfo{volume}{283}}(\bibinfo{number}{3--4}),
  \bibinfo{pages}{152--156} (\bibinfo{year}{2001}),
  \bibinfo{note}{\arxiveprint{quant-ph/0004055}}.

\bibitem[{\citenamefont{Slater}(2001{\natexlab{a}})}]{slater2001}
\bibinfo{author}{\bibfnamefont{P.~B.} \bibnamefont{Slater}},
  \emph{\bibinfo{title}{{Bures} geometry of the three-level quantum systems}},
  \bibinfo{journal}{J. Geom.\ Phys.}
  \textbf{\bibinfo{volume}{39}}(\bibinfo{number}{3}), \bibinfo{pages}{207--216}
  (\bibinfo{year}{2001}{\natexlab{a}}),
  \bibinfo{note}{\arxiveprint{quant-ph/0008069}; see also~\citep{slater2001b}}.

\bibitem[{\citenamefont{Slater}(2001{\natexlab{b}})}]{slater2001b}
\bibinfo{author}{\bibfnamefont{P.~B.} \bibnamefont{Slater}},
  \emph{\bibinfo{title}{{Bures} geometry of the three-level quantum systems.
  {II}}} (\bibinfo{year}{2001}{\natexlab{b}}),
  \bibinfo{note}{\arxiveprint{math-ph/0102032}; see also~\citep{slater2001}}.

\bibitem[{\citenamefont{Slater}(1999)}]{slater1999b}
\bibinfo{author}{\bibfnamefont{P.~B.} \bibnamefont{Slater}},
  \emph{\bibinfo{title}{{Hall} normalization constants for the {Bures} volumes
  of the {$n$}-state quantum systems}}, \bibinfo{journal}{J. Phys.\ A}
  \textbf{\bibinfo{volume}{32}}(\bibinfo{number}{47}),
  \bibinfo{pages}{8231--8246} (\bibinfo{year}{1999}),
  \bibinfo{note}{\arxiveprint{quant-ph/9904101}}.

\bibitem[{\citenamefont{Slater}(1996)}]{slater1996c}
\bibinfo{author}{\bibfnamefont{P.~B.} \bibnamefont{Slater}},
  \emph{\bibinfo{title}{Applications of quantum and classical {Fisher}
  information to two-level complex and quaternionic and three-level complex
  systems}}, \bibinfo{journal}{J. Math.\ Phys.}
  \textbf{\bibinfo{volume}{37}}(\bibinfo{number}{6}),
  \bibinfo{pages}{2682--2693} (\bibinfo{year}{1996}).

\bibitem[{\citenamefont{Arnon \etal}(1984{\natexlab{a}})\citenamefont{Arnon,
  Collins, and McCallum}}]{arnonetal1984}
\bibinfo{author}{\bibfnamefont{D.~S.} \bibnamefont{Arnon}},
  \bibinfo{author}{\bibfnamefont{G.~E.} \bibnamefont{Collins}},
  \bibnamefont{and} \bibinfo{author}{\bibfnamefont{S.}~\bibnamefont{McCallum}},
  \emph{\bibinfo{title}{Cylindrical algebraic decomposition {I}: The basic
  algorithm}}, \bibinfo{journal}{SIAM J. Comput.}
  \textbf{\bibinfo{volume}{13}}(\bibinfo{number}{4}), \bibinfo{pages}{865--877}
  (\bibinfo{year}{1984}{\natexlab{a}}), \bibinfo{note}{see
  also~\citep{arnonetal1984b}}.

\bibitem[{\citenamefont{Davenport \etal}(1987/1993)\citenamefont{Davenport,
  Siret, and Tournier}}]{davenportetal1987_t1993}
\bibinfo{author}{\bibfnamefont{J.~H.} \bibnamefont{Davenport}},
  \bibinfo{author}{\bibfnamefont{Y.}~\bibnamefont{Siret}}, \bibnamefont{and}
  \bibinfo{author}{\bibfnamefont{E.}~\bibnamefont{Tournier}},
  \emph{\bibinfo{title}{Computer Algebra: Systems and Algorithms for Algebraic
  Computation}} (\bibinfo{publisher}{Academic Press},
  \bibinfo{address}{London}, \bibinfo{year}{1987/1993}), \bibinfo{edition}{2nd}
  ed., \bibinfo{note}{transl.\ by A. Davenport and J. H. Davenport; first
  publ.\ in French 1987}.

\bibitem[{\citenamefont{Mishra}(1993)}]{mishra1993}
\bibinfo{author}{\bibfnamefont{B.}~\bibnamefont{Mishra}},
  \emph{\bibinfo{title}{Algorithmic Algebra}}
  (\bibinfo{publisher}{Springer-Verlag}, \bibinfo{address}{New York},
  \bibinfo{year}{1993}).

\bibitem[{\citenamefont{Jirstrand}(1995)}]{jirstrand1995}
\bibinfo{author}{\bibfnamefont{M.}~\bibnamefont{Jirstrand}},
  \emph{\bibinfo{title}{Cylindrical algebraic decomposition --- an
  introduction}}, \bibinfo{type}{Tech. Rep.} \bibinfo{number}{LiTH-ISY-R-1807},
  \bibinfo{institution}{Link\"oping University}, \bibinfo{address}{Link\"oping,
  Sweden} (\bibinfo{year}{1995}),
  \bibinfo{note}{\url{http://www.control.isy.liu.se/publications/doc?id=164}}.

\bibitem[{\citenamefont{Brown}(2001)}]{brown2001b}
\bibinfo{author}{\bibfnamefont{C.~W.} \bibnamefont{Brown}},
  \emph{\bibinfo{title}{Simple {CAD} construction and its applications}},
  \bibinfo{journal}{J. Symbolic Computation}
  \textbf{\bibinfo{volume}{31}}(\bibinfo{number}{5}), \bibinfo{pages}{521--547}
  (\bibinfo{year}{2001}).

\bibitem[{\citenamefont{Bloore}(1976)}]{bloore1976}
\bibinfo{author}{\bibfnamefont{F.~J.} \bibnamefont{Bloore}},
  \emph{\bibinfo{title}{Geometrical description of the convex sets of states
  for systems with spin-$\tfrac{1}{2}$ and spin-$1$}}, \bibinfo{journal}{J.
  Phys.\ A} \textbf{\bibinfo{volume}{9}}(\bibinfo{number}{12}),
  \bibinfo{pages}{2059--2067} (\bibinfo{year}{1976}).

\bibitem[{\citenamefont{Slater}(2006)}]{slater2006c}
\bibinfo{author}{\bibfnamefont{P.~B.} \bibnamefont{Slater}},
  \emph{\bibinfo{title}{Two-qubit separability probabilities and beta
  functions}} (\bibinfo{year}{2006}),
  \bibinfo{note}{\arxiveprint{quant-ph/0609006}}.

\bibitem[{\citenamefont{Arnon \etal}(1984{\natexlab{b}})\citenamefont{Arnon,
  Collins, and McCallum}}]{arnonetal1984b}
\bibinfo{author}{\bibfnamefont{D.~S.} \bibnamefont{Arnon}},
  \bibinfo{author}{\bibfnamefont{G.~E.} \bibnamefont{Collins}},
  \bibnamefont{and} \bibinfo{author}{\bibfnamefont{S.}~\bibnamefont{McCallum}},
  \emph{\bibinfo{title}{Cylindrical algebraic decomposition {II}: An adjacency
  algorithm for the plane}}, \bibinfo{journal}{SIAM J. Comput.}
  \textbf{\bibinfo{volume}{13}}(\bibinfo{number}{4}), \bibinfo{pages}{878--889}
  (\bibinfo{year}{1984}{\natexlab{b}}), \bibinfo{note}{see
  also~\citep{arnonetal1984}}.

\end{thebibliography}


\end{document}